\documentclass[12pt]{article}
\usepackage{epsf}
\title{ {\bf The Lorentz and CPT violating effects on the
$Z\rightarrow l^+ l^-$ decay }}

\author{\vspace{1cm}\\
        {\bf E. O. Iltan}
        \thanks{E-mail address:
        eiltan@heraklit.physics.metu.edu.tr}
 \\
        Physics Department, Middle East Technical University \\
        Ankara, Turkey\\}
\date{}
\begin{document}
\setlength{\baselineskip}{24pt}
\maketitle
\setlength{\baselineskip}{7mm}
\begin{abstract}
We study the Lorentz and CPT violating effects on the branching
ratio,  the CPT violating asymmetry and the ratio of the decay
width, including only the Lorentz violating effects, to the one
obtained in the standard model, for the flavor dependent part of
the lepton flavor conserving $Z\rightarrow l^+ l^-$ ($l=e,\mu,\tau
$) decay. The inclusion of the Lorentz and CPT violating effects
in the standard model contribution are too small to be detected,
since the corresponding coefficients are highly suppressed at the
low energy scale.
\end{abstract}
\thispagestyle{empty}
\newpage
\setcounter{page}{1}
\section{Introduction}
A considerable theoretical work  has been done to construct a
fundamental theory at higher scales, like the Planck scale,  which
the standard model (SM) of particle physics is its low energy
limit. In such scales,  there are hints that the Lorentz and CPT
symmetries are broken \cite{Kos1}, in contrast to their
conservation in the SM. The string theories \cite{Kos2} and the
non-commutative theories \cite{Carroll} are the examples of high
energy extensions of the SM. Even if the Lorentz and CPT symmetry
violations exist in the extended theories given above, the small
violations of these symmetries can appear at the low energy level.

The general Lorentz and CPT violating extension of the SM is
obtained in \cite{Colladay, Lehnert}. In the extension of the SM
the Lorentz and CPT violating effects are carried by the
coefficients coming from an underlying theory at the Planck scale.
These coefficients can arise from the expectation values in the
string theories or some coefficients in the non-commutative field
theories \cite{Carroll}. The loop quantum gravity \cite{Alfaro},
the space time foam \cite{Klinkhamer} and cosmological scalar
fields \cite{Bertolami} are the possible sources of the Lorentz
violating coefficients . Furthermore, the space-time varying
couplings are also associated with the Lorentz violation, and they
affect the photon dynamics \cite{KosLeh} .

In the literature, there are various studies in which some of the
coefficients are probed, by using the experiments
\cite{KTeV,Russell}. The general Lorentz and CPT violating Quantum
Electro Dynamics (QED) extension has been studied in \cite{Kos4,
Kos5} and in \cite{Kos5} the one loop renormalizability of this
extension has been shown. In \cite{EiltmuegamLrVio} the Lorentz
and CPT violating effects on the branching ratio $(BR)$ and the CP
violating asymmetry $A_{CP}$ for the  lepton flavor violating
(LFV) interactions $\mu\rightarrow e\gamma$ and $\tau\rightarrow
\mu\gamma$,  has been analyzed in the model III version of the two
Higgs doublet model (2HDM) and the relative effects of new
coefficients on these physical parameters have been studied. The
Lorentz and CPT violating effects in the Maxwell-Chern-Simons
model have been examined in \cite{Alexander, Potting} and these
effects in the non-commutative space time have been analyzed in
\cite{Bazeia}. In \cite{DColladay}, the theoretical overview of
Lorentz and CPT violation has been done; in \cite{Matthew}, the
possible signals of Lorentz violation in sensitive clock-based
experiments has been investigated and in \cite{Berger}, the
superfield realizations of Lorentz-violating extensions of the
Wess-Zumino model were presented. The threshold analysis of ultra-
high-energy cosmic rays can also be used for Lorentz and CPT
violation searches. The basis for such threshold has been
investigated in \cite{LehnertTek}.

In the present work, we study the Lorentz and CPT violating
effects on the $BR$,  the CPT violating asymmetry $(A_{CPT})$ and
the ratio $R$ of the decay width $\Gamma$, including only the
Lorentz violating effects, to the one obtained in the SM, for the
flavor dependent part of the lepton flavor conserving
$Z\rightarrow l^+ l^-$ ($l=e,\mu,\tau $) decay. The additional
contribution, coming from the Lorentz and CPT violating effects,
on the physical parameters we study is too small to be detected,
since the corresponding coefficients are highly suppressed at the
low energy scale. Our aim is to investigate the relative
importance of the coefficients which are responsible for the
Lorentz and CPT violating effects on the $BR$ of the decays under
consideration. Furthermore, we predict the possible CPT violating
asymmetry $A_{CPT}$ which is carried by the limited number of
coefficients, $e_{\mu}$ and $g_{\mu\nu\alpha}$ in the present
process. The $A_{CPT}$ is sensitive to the flavor structure of the
process, however, it is considerably small, as expected. Finally,
we study the ratio $R$ to understand contribution of the Lorentz
and CPT violating effects on the flavor structure of the decay and
we observe that these effects are too weak to be detected in the
present experiments.

The paper is organized as follows: In Section 2, we present the
theoretical expression for the decay width $\Gamma$, the $A_{CPT}$
and the ratio $R$, for the lepton flavor conserving $Z\rightarrow
l^+ l^-$ ($l=e,\mu,\tau $) decay, in the case that the Lorentz and
CPT violating effects are switched on. Section 3 is devoted to
discussion and our conclusions.
\section{The $Z\rightarrow l^+ l^-$ ($l=e,\mu,\tau $) decay with the
addition of the Lorentz and CPT violating effects }
In this section, we study the Lorentz and CPT violating effects on
the $BR$, the CPT asymmetry and the ratio $R$ for the leptonic Z
decay. In the SM,  this process is allowed at tree level and the
$BR$ is weakly sensitive to the lepton flavor. The insertion of
the Lorentz and CPT violating effects in the tree level brings new
contribution and its size is regulated by the magnitudes of the
new coefficients coming from  the tiny Lorentz and CPT violation.
The Lorentz and CPT violating lagrangian in four space-time
dimensions responsible for the decay of Z boson to the lepton pair
reads \cite{Colladay}
\begin{eqnarray}
L=\frac{i}{2} (\bar{\psi_L} \Gamma^{\mu} D_{\mu}
\psi_L+\bar{\psi_R} \Gamma^{\mu} D_{\mu} \psi_R) \label{CPTLag}
\end{eqnarray}
where
\begin{eqnarray}
\Gamma^{\mu}&=&\gamma^{\mu}+\Gamma_1^{\mu}\, , \nonumber \\
\Gamma_1^{\mu}&=&c^{\alpha\mu}\,\gamma_{\alpha}+d^{\alpha\mu}\,\gamma_5\,
\gamma_{\alpha}+e^{\mu}+i
f^{\mu}\,\gamma_5+\frac{1}{2}\,g^{\lambda\nu\mu}\,
\sigma_{\lambda\nu}  \, . \label{GammaM2}
\end{eqnarray}
Here the coefficients $c_{\alpha \mu}$, $d_{\alpha\mu}$,
$e_{\mu}$, $f_{\mu}$ and $g_{\lambda\nu\mu}$ are responsible for
the Lorentz violation. Even if the U(1) charge symmetry and
renormalizability does not exclude the part of lagrangian
including the coefficients $e_{\mu}$, $f_{\mu}$ and
$g_{\lambda\nu\mu}$, they are not compatible with the electroweak
structure of the SM extension. However, the possible
nonrenormalizable higher dimensional operators respecting the
electroweak symmetry and including Higgs field with vacuum
expectation value can create these highly suppressed terms (see
\cite{Colladay} for details). In our analysis we also take these
terms into consideration since they are sources of the CPT
violation (\cite{Kos5}).

Now, we would like to present the additional vertex due to the
Lorentz and CPT violating effects for the $Z\rightarrow l^+ l^-$
decay:
\begin{eqnarray}
V_{LorVio}=\frac{-i\,Q_l\,e}{s_W\, c_W}\, \Bigg\{
c^{\alpha\mu}\,\gamma_{\alpha}+d^{\alpha\mu}\,\gamma_5\,
\gamma_{\alpha}+e^{\mu}+i \,
f^{\mu}\,\gamma_5+\frac{1}{2}\,g^{\lambda\nu\mu}\,
\sigma_{\lambda\nu}  \Bigg \}\, (c_L^l\, L+c_R^l\,R)
 \, ,
\label{LorCPTVio}
\end{eqnarray}
where $L(R)=\frac{1}{2}(1\pm\gamma_5)$,
$c_L^l=\frac{-1}{2}+s_W^2$, $c_R^l=s_W^2$ and $Q_l=-1$. Our aim is
to calculate the decay width of the $Z\rightarrow l^+ l^-$ process
including the Lorentz violating effects. It is known that the
invariant phase-space elements in the presence of Lorentz
violation are modified \cite{Potting}. In the conventional case
where there are no Lorentz violating effects, the well known
expression for decay width in the $Z$ boson rest frame reads
\begin{eqnarray}
d\Gamma&=&\frac{(2\pi)^4}{6\,m_Z}\, \delta^{(4)}(p_Z-q_1-q_2)\,
\frac{d^3 q_1}{(2\pi)^3\,2\,E_1}\,\frac{d^3 q_2}{(2\pi)^3\,2\,E_2}
 \nonumber \\ &\times& |M|^2 (p_Z,q_1,q_2)
\label{DecWid}
\end{eqnarray}
with the four momentum vector of Z boson (lepton, anti-lepton)
$p_Z$ ($q_1, \, q_2$), and the matrix element $M$ for the process
$Z\rightarrow l^+\, l^-$.  The inclusion of the new Lorentz
violating  parameters changes the lepton dispersion relation and
an additional part in the phase space element $\frac{d^3
q_i}{(2\pi)^3\,2\,E_i}$ is switched on. The variational procedure
generates the Dirac equation
\footnote{In the case of the existence of the new Lorentz
violating effects lying in the part $-\bar{\psi} M \psi$ where
$M=m+M_1$, $M_1=a_\mu \gamma_\mu+b_\mu \gamma_5
\gamma_\mu+\frac{1}{2} H_{\mu\nu} \sigma_{\mu\nu}$ (see
\cite{Kos5} for details), the modified Dirac equation becomes
$(\gamma^{\mu} q_{\mu}-m-M_1+\Gamma_1^{\mu} q_{\mu})\, \psi=0$}.
\begin{equation}
(\gamma^{\mu} q_{\mu}-m+\Gamma_1^{\mu} q_{\mu})\, \psi=0 \nonumber
\, .
\end{equation}
and a small modification on $E_i$ in the phase space element is
obtained. In our case, the corresponding dispersion relation is a
complicated function of the various Lorentz violating parameters
(see \cite{Lehnert} for example). In addition to this, the crowd
of Lorentz violating parameters causes large number of fixed
directions and makes the angular integrations complicated, since
the amplitude has a functional dependence of these angular
variables. Finally, spin sums in the final state are not trivial
in the case of Lorentz violating effects since the phase factors
depend on the outgoing lepton polarizations (see \cite{Altschul}
for details). Therefore, in the present work, we do not take into
account these tiny additional effects \footnote{ The modified
Dirac equation for the outgoing lepton in the present case is
$(\gamma^{\mu} q_{\mu}-m+\Gamma_1^{\mu} q_{\mu})\, \psi=0$.
Now, we assume that all the Lorentz violating coefficients, except
$c^{00}, d^{00}, e^{0}, f^{0}$ and $g^{ijk}$, are vanishing.
Furthermore, $g^{ijk}$ is small compared to other coefficients.
After some algebra, the dispersion relation is obtained as \\
$(q^2-m_l^2+2\, m_l\, E\, e^0)^2-4\, m_l^2\, E^2\, (s^{00})^2=0$,
where $(s^{00})^2=(c^{00})^2-(d^{00})^2- (f^{0})^2$ and the energy
eigenvalues read $E^n_{\pm}\simeq -m_l\,(e^0+(-1)^n\, s^{00}) \pm
\sqrt{\vec{q}^2+m_l^2\,\Big(1+ (e^0+s^{00})^2\Big)}$ where $n=1$
or $2$. Following the integration over the anti lepton
four momentum $q_2$, the phase factor $1/E_1$ is replaced by \\
$1/\Big( -m_l\,(e^0+s^{00})+ \sqrt{\vec{q}^2+m_l^2\,\Big(1+
(e^0+s^{00})^2\Big)}\,\, \Big) \simeq
\Big(1+m_l\,(e^0+s^{00})/\sqrt{\vec{q}_1^2+m_l^2}\Big)/
\sqrt{\vec{q}_1^2+m_l^2}$. We expect that the factor
$1/\sqrt{\vec{q}_1^2+m_l^2}$ in the additional part further
suppresses the Lorentz violating effects in the phase factor,
after the kinematical integration over the lepton four momentum
$q_1$.} and use the conventional expression for the decay width in
the $Z$ boson rest frame (see eq. (\ref{DecWid})).
With the inclusion of the Lorentz violating effects in the matrix
element, the Lorentz violating part of the decay width
$\Gamma(Z\rightarrow l^+ l^-)$ is obtained as
\begin{eqnarray}
\Gamma_{LorVio}&=&\frac{e^2\, Q_l^2}{48\,\pi\,m_Z^2\, c_W^2\,
s_W^2}\, s_l\, \{  c_{00} \, \Big( 2\, m_l^2- m_Z^2\,
(1-4\,s_W^2+8\,s_W^4) \Big) \nonumber \\ &+& d_{00} \, (2\,
m_l^2-m_Z^2)\,(1-4\,s_W^2)+\frac{1}{2}\,g\,m_Z\,m_l\,
(1-4\,s_W^2)\,
 \} \, . \label{GamLorVio}
\end{eqnarray}
Here $s_l=\sqrt{\frac{m_Z^2}{4}-m_l^2}$, the parameters $c_{00}$
and $d_{00}$ are the zeroth components of the coefficients
$c_{\alpha\beta}$ and $d_{\alpha\beta}$ and the last term $g$ is
$g=\epsilon_{ijk}\, g^{ikj}$ , where $i,j,k=1,2,3$.
Notice that we take only the additional part of the decay width
which is linear in the Lorentz violating coefficients. The eq.
(\ref{GamLorVio}) shows that $\Gamma_{LorVio}$ depends on the CPT
even ($c_{00}$ and $d_{00}$) and the CPT odd $g$ coefficients.
Using $\Gamma_{LorVio}$ it is easy to calculate the Lorentz
violating part of the $BR$ ($BR_{LorVio}$) as:
\begin{eqnarray}
BR_{LorVio}=\frac{\Gamma_{LorVio}}{\Gamma_{Z}} \, , \label{BRDef}
\end{eqnarray}
where the $\Gamma_{Z}$ is the total decay width of the $Z$ boson
and its numerical value is $\Gamma_Z=2.490\, (GeV)$.

The coefficient $g$ switches on the CPT asymmetry and it reads
\begin{eqnarray}
A_{CPT}&=&\frac{(1-4\,s_W^2)\,m_l\,g}{D} \, , \label{ACPT1}
\end{eqnarray}
where
\begin{eqnarray}
D&=&2\,m_Z\, \Bigg( \Big(1-4\,s_W^2\,
(1-2\,s_W^2)\Big)-\frac{m_l^2}{m_Z^2}\,\Big(1+8\,s_W^2\,(1-2\,s_W^2)
\Big) \nonumber
\\ &+& \Big(2\,\frac{m_l^2}{m_Z^2}-(1-4\,s_W^2(1-2\,s_W^2))
\Big)\,c_{00}+(1-4\,s_W^2)\,(2\,\frac{m_l^2}{m_Z^2}-1)\, d_{00}
\Bigg)\, . \label{ACPTD}
\end{eqnarray}
This equation shows that the $A_{CPT}$ depends on the flavor part
of the decay under consideration and becomes larger for the
heavier lepton pair decay.

Finally, we study the ratio
$R=\frac{\Gamma^{flavor}_{LorVio}}{\Gamma^{flavor}_{SM}}$
\begin{eqnarray}
R=\frac{4\,m_l\, c_{00}+(1-4\,s_W^2)\, (4\,d_{00}\,
m_l+g\,m_Z)}{2\,m_l\,(1+8\,s_W^2\,(1-2\,s_W^2))} \label{Rml}
\end{eqnarray}
where $\Gamma^{flavor}_{LorVio}$ ($\Gamma^{flavor}_{SM}$) is the
flavor dependent part of the decay width including only the
Lorentz violating (the SM without Lorentz violating) effects. This
ratio is sensitive to the lepton mass and it is dominant for the
light lepton pair decay.
\section{Discussion}
The SM is invariant under the Lorentz and CPT transformations,
however, the small violations of Lorentz and CPT symmetry,
possibly coming from an underlying theory at the Planck scale, can
arise in the extensions of the SM. In this section, we analyze the
Lorentz and CPT violating  effects on the $BR$ and the  $A_{CPT}$
for the $Z\rightarrow l^+ l^-$ ($l=e,\mu, \tau$) decays, in the SM
extension. Furthermore, we study the ratio
$R=\frac{\Gamma^{flavor}_{LorVio}}{\Gamma^{flavor}_{SM}}$ to
understand contribution of the Lorentz and CPT violating effects
on the flavor structure of the decay. It is well known that these
effects are tiny to be observed, however, it would be interesting
to see the relative behaviors of different coefficients, which are
responsible for the violation of the Lorentz and CPT symmetry.

The natural suppression scale for these coefficients can be taken
as the ratio of the light one $m_l$ to the one of the order of the
Planck mass. Therefore, the coefficients which carry the Lorentz
and CPT violating effects are roughly in the range of
$10^{-23}-10^{-17}$ \cite{Russell}. Here the first (second) number
represent the electron mass $m_e$ ($m_{EW}\sim 250\,GeV$) scale.
We take the numerical values of the coefficients $|d|, |c|,|e|,
|g|$ at the order of the magnitude of $10^{-23}-10^{-17}$.

\begin{figure}[htb]
\vskip -3.0truein \centering \epsfxsize=6.8in
\leavevmode\epsffile{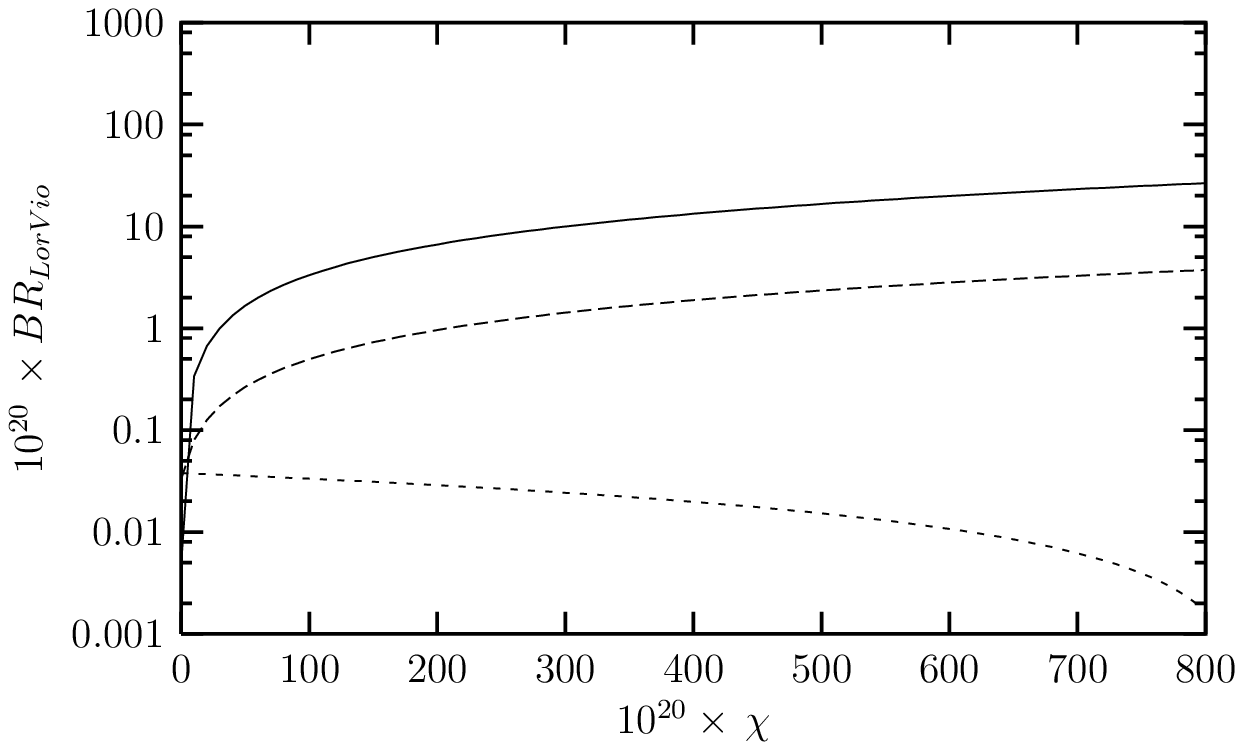} \vskip -3.0truein \caption[]{
The magnitude of the coefficient dependence of $BR_{LorVio}$ for
the decay $Z\rightarrow \tau^+\tau^-$. Here solid (dashed, small
dashed) line represents the dependence to the coefficient $c_{00}$
($d_{00}$, $g$), in the case that the other coefficients have the
same numerical value $10^{-20}$. } \label{BRZtautau}
\end{figure}

\begin{figure}[htb]
\vskip -3.0truein \centering \epsfxsize=6.8in
\leavevmode\epsffile{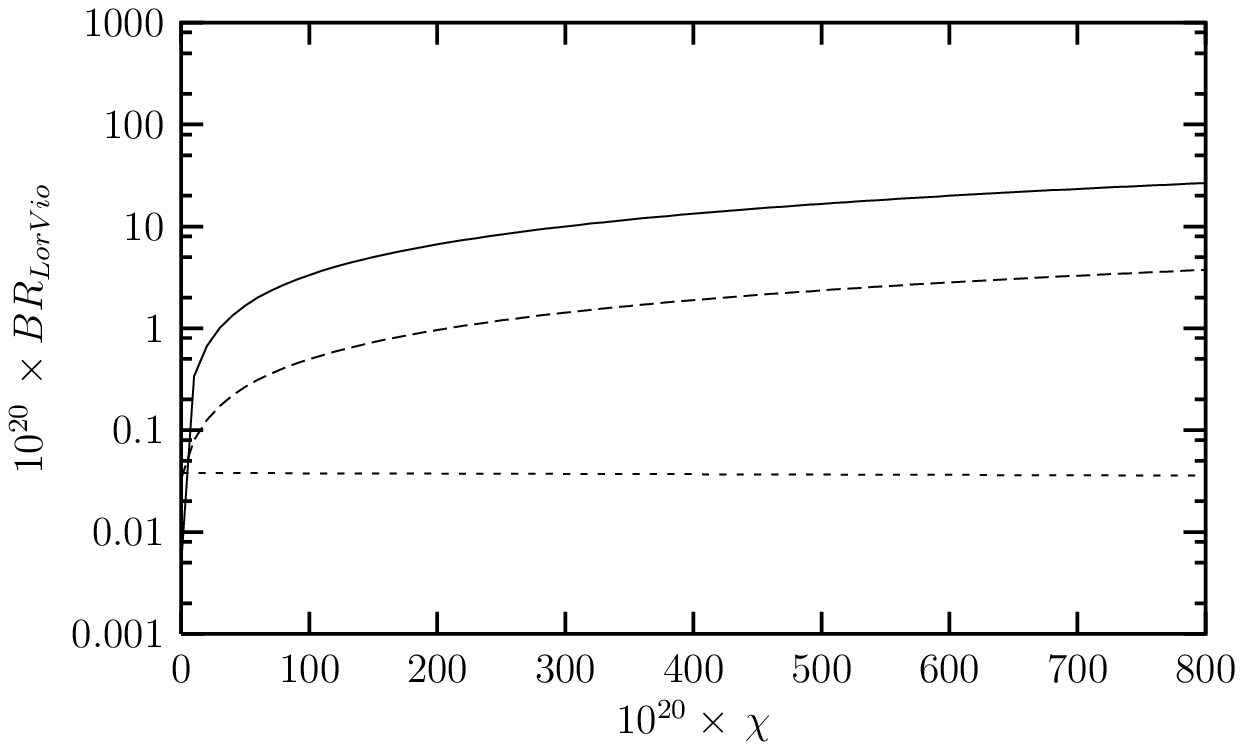} \vskip -3.0truein \caption[]{The
same as Fig. \ref{BRZtautau} but for the decay $Z\rightarrow \mu^+
\mu^-$.} \label{BRZmumu}
\end{figure}

In Fig. \ref{BRZtautau} (\ref{BRZmumu}), we present the magnitude
of the coefficient dependence of the Lorentz violating part of the
$BR$ ($BR_{LorVio}$) for the decay $Z\rightarrow \tau^+\tau^- \,
(\mu^+\mu^-)$. Here solid (dashed, small dashed) line represents
the dependence to the coefficient $c_{00}$ ($d_{00}$, $g$), in the
case that the other coefficients have the same numerical value
$10^{-20}$. Notice that, in the figures, the parameter $\chi$
denotes to the size of $c_{00}$, $d_{00}$ and $g$ for different
lines. It is observed that the $BR$ is more sensitive to the
coefficient $c_{00}$ compared to the others and the contribution
of the new effects to the $BR$ is at the order of the magnitude of
$10^{-19}$ for the large values of the coefficient $c_{00}\sim
10^{-17}$. With the increasing values of the coefficient $g$ the
$BR$ decreases for $Z\rightarrow \tau^+\tau^-$ decay and this
effect is weak for the light lepton pair case, namely
$Z\rightarrow \mu^+\mu^-$ decay. Notice that the Lorentz violating
coefficient dependence of the $BR_{LorVio}$ for the decay
$Z\rightarrow e^+ e^- $ is almost the same as the one for the
decay $Z\rightarrow \mu^+\mu^-$.

\begin{figure}[htb]
\vskip -3.0truein \centering \epsfxsize=6.8in
\leavevmode\epsffile{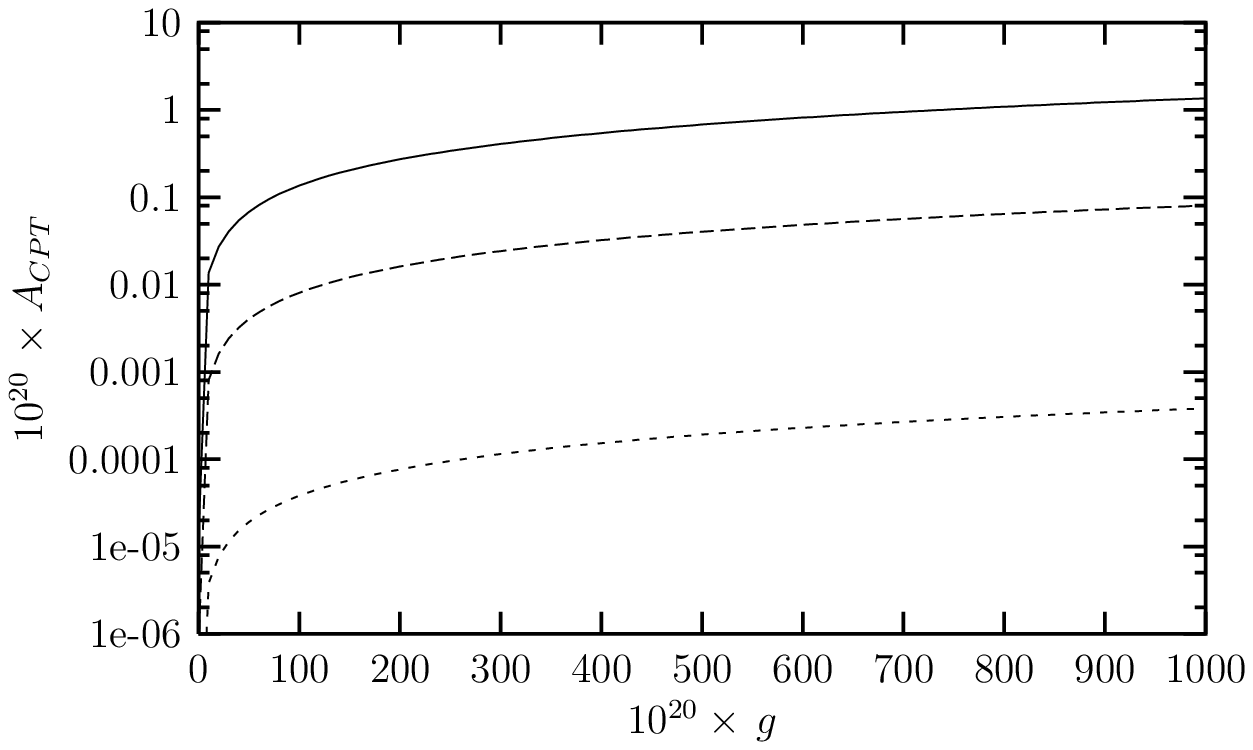} \vskip -3.0truein \caption[]{The
magnitude of the coefficient dependence of the $A_{CPT}$ for the
decay $Z\rightarrow l^+ l^-$, ($l=e,\mu,\tau$) Here solid (dashed,
small dashed) line represents the $A_{CPT}$ for the decay
$Z\rightarrow \tau^+\tau^-$ ($\mu^+ \mu^-, \, e^+ e^- $).}
\label{ACPTZ}
\end{figure}

Now, we analyze the CPT violating asymmetry $A_{CPT}$ for the
decays under consideration. The coefficient $g$ and the lepton
flavor are responsible for this violation as seen in the eq.
(\ref{ACPT1}). Fig. \ref{ACPTZ} is devoted to the magnitude of the
coefficient dependence of the $A_{CPT}$. Here solid (dashed, small
dashed) line represents the $A_{CPT}$ for the decay $Z\rightarrow
\tau^+\tau^-$ ($\mu^+\mu^-, \, e^+ e^- $). The $A_{CPT}$ is
sensitive to the lepton flavor and it is at the order of the
magnitude of $10^{-20}$ ($10^{-21}$, $10^{-23}$) for the large
values of the coefficient $g\sim 10^{-17}$, for the decay
$Z\rightarrow \tau^+\tau^- \,(\mu^+\mu^-, \, e^+ e^- )$. The
$A_{CPT}$ enhances with the increasing lepton mass.

\begin{figure}[htb]
\vskip -3.0truein \centering \epsfxsize=6.8in
\leavevmode\epsffile{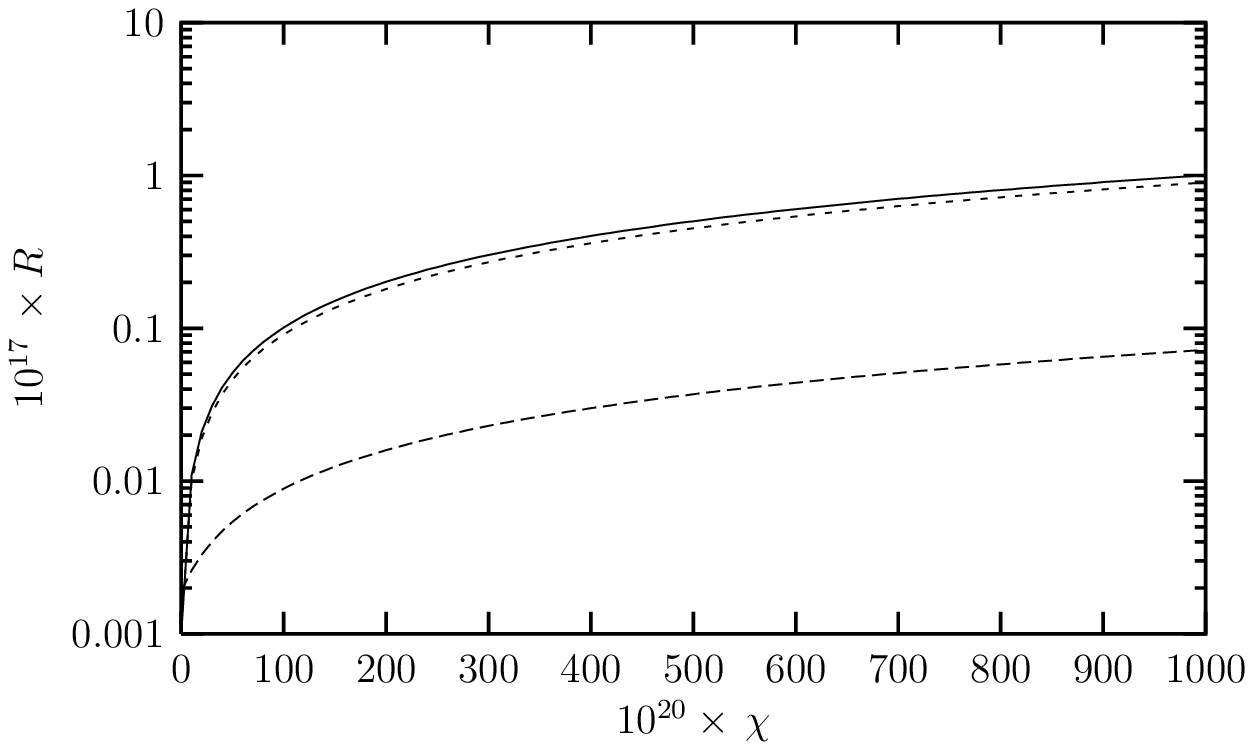} \vskip -3.0truein \caption[]{The
magnitude of the coefficient dependence of the ratio R for the
decay $Z\rightarrow \tau^+ \tau^-$. Here solid (dashed, small
dashed) line represents the dependence to the coefficient $c_{00}$
($d_{00}$, $g$), in the case that the other coefficients have the
same numerical value $10^{-20}$.} \label{RZtautau}
\end{figure}

\begin{figure}[htb]
\vskip -3.0truein \centering \epsfxsize=6.8in
\leavevmode\epsffile{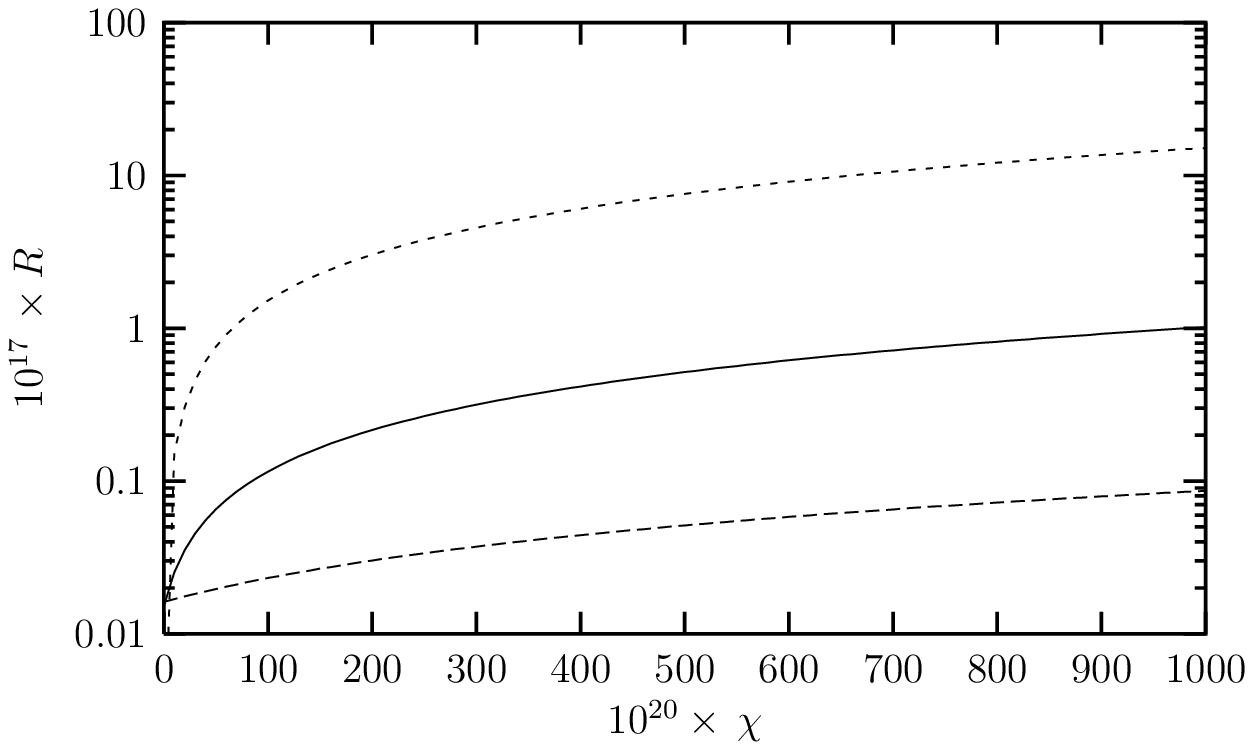} \vskip -3.0truein \caption[]{The
same as Fig. \ref{RZtautau} but for the decay $Z\rightarrow \mu^+
\mu^-$}. \label{RZmumu}
\end{figure}

\begin{figure}[htb]
\vskip -3.0truein \centering \epsfxsize=6.8in
\leavevmode\epsffile{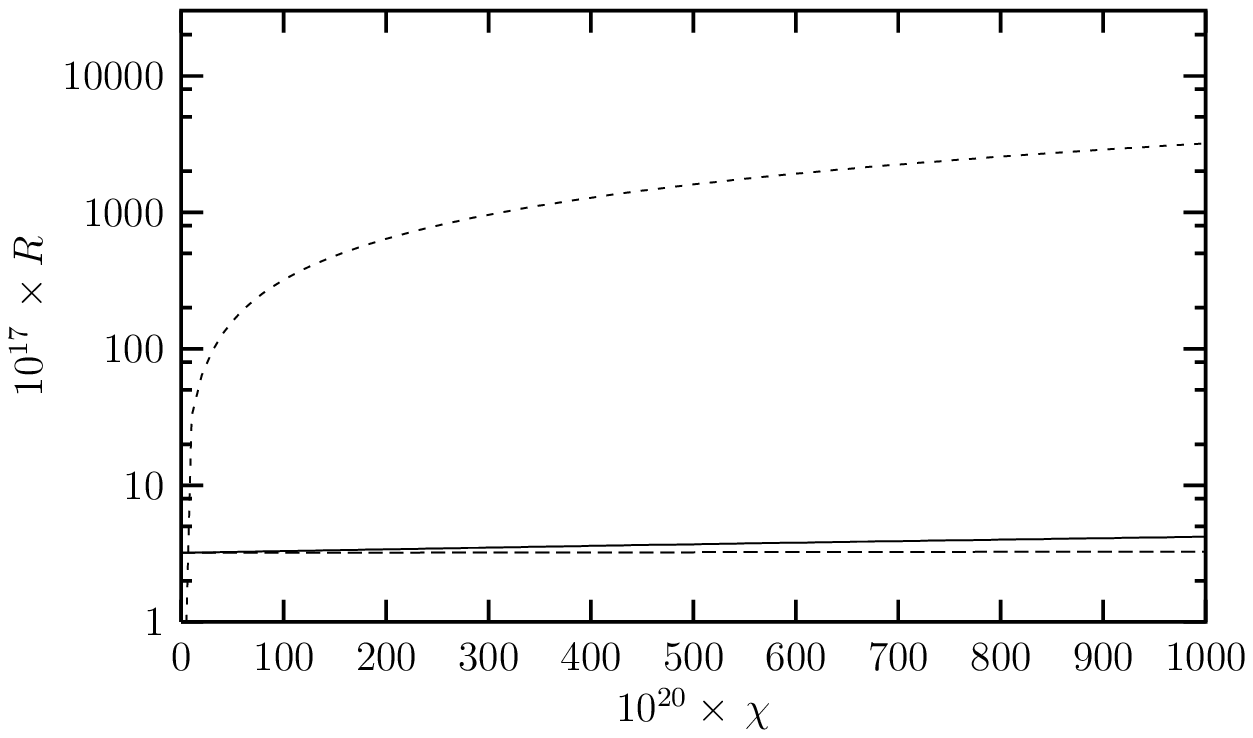} \vskip -3.0truein \caption[]{The
same as Fig. \ref{RZtautau} but for the decay $Z\rightarrow e^+
e^-$. } \label{RZee}
\end{figure}

Finally, we study the ratio
$R=\frac{\Gamma^{flavor}_{LorVio}}{\Gamma^{flavor}_{SM}}$ and we
present the the magnitude of the coefficient dependence of the
ratio R for the decay $Z\rightarrow \tau^+ \tau^- \, (\mu^+ \mu^-,
\, e^+ e^- )$ in Fig. \ref{RZtautau} (\ref{RZmumu}, \ref{RZee}).
Here solid (dashed, small dashed) line represents the dependence
to the coefficient $c_{00}$ ($d_{00}$, $g$), in the case that the
other coefficients have the same numerical value $10^{-20}$. The
ratio R is more sensitive to the coefficients $c_{00}$ and $g$
compared to the coefficient $d_{00}$ and it is at the order of the
magnitude of $10^{-17}$ for the large values of the coefficient
$c_{00}\,(g)\sim 10^{-17}$, for the decay $Z\rightarrow
\tau^+\tau^-$. For the $Z\rightarrow \mu^+ \mu^-$ decay, R is more
sensitive to the coefficient $g$ and it can reach the values of
$10^{-16}$. In the case of $Z\rightarrow e^+ e^-$ decay, the
sensitivity of R to the coefficients $c_{00}$ and $d_{00}$ is
weak, however, it enhances up to the values of $10^{-14}$.

At this stage we would like to summarize our results:

We analyze the Lorentz and CPT violating  effects on the $BR$,
$A_{CPT}$ and the ratio R and we study the relative behaviors of
different coefficients by taking their numerical values at the
order of the magnitude of $10^{-20}-10^{-17}$:
\begin{itemize}
\item The contribution of the Lorentz and CPT violating part to
the $BR$ of the decays $Z\rightarrow l^+ l^-$ ($l=e,\mu, \tau$) is
at most at the order of the magnitude of $10^{-19}$ for the large
values of the coefficients $c_{00}$, $d_{00}$ and $g$ and these
numbers are too small to be detected.

\item We predict the numerical value of $A_{CPT}$ at the order of
$10^{-20}$ ($10^{-21}$, $10^{-23}$) for the large values of the
coefficient $g\sim 10^{-17}$ for the decay $Z\rightarrow
\tau^+\tau^- \, (\mu^+ \mu^-, \,e^+\,e^-)$. This physical
parameter is driven by the coefficient $g$ and the lepton flavor.
It enhances with the increasing values of lepton mass.

\item We study the ratio
$R=\frac{\Gamma^{flavor}_{LorVio}}{\Gamma^{flavor}_{SM}}$ and we
observe that its sensitivity to the coefficient $g$ ($c_{00}$,
$d_{00}$) increases (decreases) with the decreasing values of the
lepton mass. It is at the order of the magnitude of $10^{-17}$
($10^{-16}$, $10^{-14}$) for $g \sim 10^{-17}$, for the decay
$Z\rightarrow \tau^+\tau^-$ ($Z\rightarrow \mu^+ \mu^-$,
$Z\rightarrow e^+\,e^-$).
\end{itemize}

\section{Acknowledgement}
This work has been supported by the Turkish Academy of Sciences in
the framework of the Young Scientist Award Program.
(EOI-TUBA-GEBIP/2001-1-8)
\newpage

\begin{thebibliography}{1}
%
\bibitem{Kos1} V. A. Kostelecky, ed. ,CPT and Lorentz Symmetry,
World Scientific,Singapore,1999.
%
\bibitem{Kos2} V. A. Kostelecky and S. Samuel, {\it Phys. Rev}
{\bf D39}, 683 (1989); {\it Phys. Rev} {\bf D40}, 1886 (1989);
{\it Phys. Rev. Lett.} {\bf 63}, 224 (1989); {\it Phys. Rev.
Lett.} {\bf 66}, 1811 (1991); {\it Phys. Lett.} {\bf B381}, 89
(1996); {\it Phys. Rev} {\bf D63}, 046007 (2001); V. A.
Kostelecky, M. Perry and R. Potting, {\it Phys. Rev. Lett.} {\bf
D84}, 4541 (2000).
%
\bibitem{Carroll} S. M. Carroll, et. al, {\it Phys. Rev. Lett.} {\bf 87},
141601 (2001); Z. Guralnik, et. al. {\it Phys. Lett.} {\bf B517},
450 (2001); A. Anisimov et.al, {\it Phys. Rev} {\bf D65}, 085032
(2002), C. E. Carlson, et.al., {\it Phys. Lett.} {\bf B 518}, 201
(2001).
%
\bibitem{Colladay} D. Colladay and V. A. Kostelecky, {\it Phys. Rev}
{\bf D55}, 6760 (1997); {\bf D58}, 116002 (1998); {\it Phys.
Lett.} {\bf B 511}, 209 (2001);
%
\bibitem{Lehnert} V. A. Kostelecky and R. Lehnert, {\it Phys. Rev.}
{\bf D 63}, 065008 (2002).
%

\bibitem{Alfaro} Alfaro et al, {\it Phys. Rev. Lett.}
{\bf 84}, 2318 (2000), {\it Phys. Rev} {\bf D65}, 103509 (2002).
%
\bibitem{Klinkhamer} F. R. Klinkhamer and C. Rupp, {\it Phys. Rev}
{\bf D70}, 045020 (2004).
%
\bibitem{Bertolami} V. A. Kostelecky et. al., {\it Phys. Rev} {\bf D68},
123511 (2003); Bertolami et al, {\it Phys. Rev} {\bf D69}, 083513
(2004).
%
\bibitem{KosLeh} V. A. Kostelecky, R. Lehnert, and Perry, {\it Phys. Rev}
{\bf D68}, 123511 (2003).
%
\bibitem{KTeV} KTeV Collaboration, Y. B. Hsiung et.al.,
{\it Nucl. Phys. Proc. Suppl.} {\bf 86}, 312 (2000); Opal
Collaboration, R. Ackerstaff et.al., {\it Z. Phys.} {\bf C76}, 401
(1997); DELPHI Collaboration, M. Feindt et. al., prep. DELPHI
97-98 CONF 80, (1997), BELLE Collaboration, K. Abe  et. al., {\it
Phys. Rev Lett. } {\bf 86}, 3228 (2001); V. A. Kostelecky, and R.
Potting, {\it Phys. Rev.} {\bf D51}, 3923 (1995); D. Colladay and V.
A. Kostelecky, {\it Phys. Lett.} {\bf B344}, 259 (1995); {\it
Phys. Rev.} {\bf D 52}, 6224 (1995); V. A. Kostelecky, and R. Van
Kooten, {\it Phys. Rev. } {\bf D54} 5585 (1996); V. A. Kostelecky,
{\it Phys. Rev. Lett.} {\bf 80} 1818 (1998); {\it Phys. Rev.} {\bf
D61} 016002 (2000); {\bf D64} 076001 (2001); N. Isgur  et. al,
{\it Phys. Lett. } {\bf B515}, 333 (2001); O. Bertolami et.al,
{\it Phys. Lett. } {\bf B395}, 178 (1997); D. Bear et. al., {\it
Phys. Rev. Lett.} {\bf 85}, 5038 (2000); D. F. Philllips et. al.,
{\it Phys. Rev.} {\bf D63}, 111101 (2001); M. A. Humphrey et. al.,
physics/0103068; {\it Phys. Rev.} {\bf A 62}, 063405 (2000); V. A.
Kostelecky, and C. D. Lane, {\it Phys. Rev} {\bf D60} 116010
(1999); {\it J. Math. Phys.} {\bf 40} 6245 (1999); R. Bluhm
et.al., {\it Phys. Rev. Lett.} {\bf 88} 090801 (2002).H.  Dehmelt
et.al.,  {\it Phys. Rev. Lett.} {\bf 83}, 4694 (1999); R.
Mittleman et.al., {\it Phys. Rev. Lett.} {\bf 83}, 2116 (1999); G.
Gabrielse et.al., {\it Phys. Rev. Lett.} {\bf 82}, 3198 (1999); R.
Bluhm et al., {\it Phys. Rev. Lett.} {\bf 82}, 2254 (1999); {\it
Phys. Rev. Lett.} {\bf 79}, 1432 (1997); R. Bluhm and V. A.
Kostelecky, {\it Phys. Lett.} {\bf 84}, 1381 (2000); V. A.
Kostelecky and M. Mewes, {\it Phys. Rev. Lett.} {\bf 87}, 251304
(2001); V. W. Hughes et.al, {\it Phys. Rev. Lett.} {\bf 87},
111804 (2001); R. Bluhm et.al., {\it Phys. Rev. Lett.} {\bf 84},
1098 (2000),
%
\bibitem{Russell} R. Bluhm, V. A. Kostelecky and N. Russell,
{\it Phys. Rev.} {\bf D57}, 3932 (1998)
%
\bibitem{Kos4} V. A. Kostelecky and A. G. M. Pickering,
{\it Phys. Rev. Lett} {\bf 91}, 031801 (2003).
%
\bibitem{Kos5} V. A. Kostelecky, C. D. Lane and A. G. M. Pickering,
{\it Phys. Rev} {\bf D65} 056006 (2002).
%
\bibitem{EiltmuegamLrVio} E. O. Iltan,
{\it JHEP} {\bf 0306} 016 (2003).
%
\bibitem{Alexander} S. M. Carroll, G. B. Field and R. Jackiw,
{\it Phys. Rev.} {\bf D41}, 1231 (1990); M. Perez-Victoria, {\it
JHEP} {\bf 032}, 0104 (2001); L. Cervi et al, {\it Phys. Rev.} {\bf
64}, 105003, (2001); Alexander A. Andrianov, P. Giacconi, R.
Soldati, {\it JHEP} {\bf 0202} 030 (2002), H. Belich, M.M.
Ferreira, J.A. Helayel-Neto, M.T.D. Orlando , {\it Phys. Rev} {\bf
D67} 125011 (2003), H. Belich, J.L. Boldo, L.P. Colatto, J.A.
Helayel-Neto, A.L.M.A. Nogueira; {\it Phys. Rev} {\bf D68} 065030
(2003); A.P. Baeta Scarpelli, H. Belich, J.L. Boldo, L.P. Colatto,
J.A. Helayel-Neto, A.L.M.A. Nogueira, hep-th/0305089.
%
\bibitem{Potting} R. Lehnert and R. Potting, {\it Phys. Rev. Lett.}
{\bf 93}, 110402, (2004); hep-ph/0408285
%
\bibitem{Bazeia} D. Bazeia, T. Mariz, J.R. Nascimento, E. Passos, R.F.
Ribeiro, {\it J.Phys.} {\bf A36} 4937 (2003)
%
\bibitem{DColladay} D. Colladay, AIP Conf.Proc. 672 (2003) 65, hep-ph/0301223.
%
\bibitem{Matthew} M. Mewes, hep-ph/0307161, R.Bluhm, V. A. Kostelecky,
C. Lane, N. Russell, {\it Phys. Rev} {\bf D68} 125008 (2003).
%
\bibitem{Berger} M. S. Berger, {\it Phys. Rev} {\bf D68} 115005
(2003).
%
\bibitem{LehnertTek} R. Lehnert, {\it Phys. Rev} {\bf D68} 085003
(2003).
%
\bibitem{Altschul} B. Altschul, {\it Phys. Rev} {\bf D70} 056005
(2004).
%
\end{thebibliography}
\end{document}